\documentclass[twocolumn,tighten]{aastex63}
\usepackage{amsmath,amstext}
\usepackage[T1]{fontenc}
\usepackage{ae,aecompl}
\usepackage[utf8]{inputenc}
\usepackage[figure,figure]{hypcap}
\usepackage{enumitem}
\usepackage{bm}

\newcommand{\mrk}{Mrk~421}
\newcommand{\emu}{$\times 10^{-3}$ cm$^{-6}$ pc}
\defcitealias{2021ApJ...918...83D}{D21}

\shorttitle{Hot CGM in emission}
\shortauthors{Bhattacharyya et al.}

\begin{document}

\title{The hot circumgalactic medium of the Milky-Way: new insights from XMM-Newton observations}

\correspondingauthor{Souradip Bhattacharyya}
\email{bhattacharyya.37@osu.edu}

\author[0000-0001-6442-5786]{Souradip~Bhattacharyya}
\affiliation{The Ohio State University, Department of Astronomy, Columbus, OH 43210, USA}
\affiliation{Center for Cosmology and Astroparticle Physics, 191 West Woodruff Avenue, Columbus, OH 43210, USA}

\author{Sanskriti~Das}
\affiliation{The Ohio State University, Department of Astronomy, Columbus, OH 43210, USA}
\affiliation{Center for Cosmology and Astroparticle Physics, 191 West Woodruff Avenue, Columbus, OH 43210, USA}

\author{Anjali~Gupta}
\affiliation{Columbus State Community College, 550 E Spring St., Columbus, OH 43215, USA}

\author{Smita~Mathur}
\affiliation{The Ohio State University, Department of Astronomy, Columbus, OH 43210, USA}
\affiliation{Center for Cosmology and Astroparticle Physics, 191 West Woodruff Avenue, Columbus, OH 43210, USA}

\author{Yair Krongold}
\affiliation{Instituto de Astronomia, Universidad Nacional Autonoma de Mexico, 04510 Mexico City, Mexico}

\begin{abstract}

We present XMM-Newton observations around the sightline of \mrk{}.  The emission spectrum of the Milky Way circumgalactic medium (CGM) shows that a two phase model is a better fit to the data compared to a single phase model; in addition to the warm-hot virial phase at  log ($T/$K) = $6.33_{-0.02}^{+0.03}$ , a hot super-virial phase at log ($T/$K) = $6.88_{-0.07}^{+0.08}$ is required. Furthermore, we present  observations of five fields within $5$ degrees of the primary field. Their spectra also require a two-phase model at warm-hot and hot temperatures.  The hot phase, first discovered in Das et al. 2019, appears to be widespread. By chemical tagging we show that emission from the supevirial phase comes from the L-shell transitions of Fe XVIII--FeXXII, and that the range of temperatures probed in emission is distinct from that in absorption. We detect scatter in temperature and emission measure (EM) in both the phases, and deduce that there is small-scale density inhomogeneity in the MW CGM. The emitting gas likely has higher density, possibly from regions close to the disk of the MW, while the absorption in the virial phase may arise from low-density gas extended out to the virial radius of the MW.  The presence of the super-virial phase far from the regions around the Galactic center implicates physical processes unrelated to the activity at the Galactic center. Hot outflows resulting from star-formation activity throughout the Galactic disk are likely responsible for producing this phase.

\end{abstract}

\keywords{Circumgalactic medium–X-ray astronomy-Milky Way galaxy}

\section{Introduction} \label{sec:intro}

The circumgalactic medium (CGM) is a diffuse gaseous halo that surrounds the disk of spiral galaxies, extending upto and beyond the virial radius, and holds a large fraction of the baryon content of these galaxies. The properties of the CGM are determined by their host dark matter halo as well as the combined processes of accretion of cold, pristine gas from the intergalactic medium (IGM) and feedback of hot, metal enriched gas from the galaxy \citep{2017ARA&A..55..389T}. The gas contained in the CGM has been posited to explain the 'missing baryon' problem \citep{2012ApJ...756L...8G}.

The CGM, through studies at different energies, has been found to be made up of multiple components, each corresponding to a certain temperature and density, much like the composition of the interstellar medium (ISM) \citep{2011piim.book.....D}. Particularly,
the $z=0$ absorption and emission lines of highly ionized metal ions along extragalactic sight lines has been studied to characterize the extreme phase(s) of this gas. For example, the O VII K$\alpha$ transition at 0.574 keV is dominant in the phase of the CGM with temperatures of $\sim 10^6$ K. Therefore, the strength of the respective emission or absorption line in the soft X-ray spectrum can be used to characterize the $\sim 10^6$ K phase of the CGM. 

This particular phase had been linked to the coronal gas of the Milky-Way (MW) that should exist at the virial temperature of the dark matter halo \citep{1956ApJ...124...20S}. The detection of a component with temperatures of $\sim 10^6$ K ($ \sim 0.2$ keV) in emission \citep{2000ApJS..128..171S,2000ApJ...543..195K,2009ApJ...690..143Y,2009PASJ...61..805Y,2010ApJ...723..935H,2013ApJ...773...92H,2015ApJ...799..117H,2018ApJ...862...34N} and absorption \citep{2012ApJ...756L...8G,2016ApJ...828L..12N,2017ApJ...836..243G,2018MNRAS.474..696G} has fulfilled this prediction and consequently it has been named as the the virial or the warm-hot component of the CGM. 

Detection of another component at even higher temperatures of $ > 10^7$ K ($ > 0.5$ keV), has recently been confirmed. Beginning with the sight line of the blazar 1ES 1553+113, this component was revealed in the MW's CGM both in absorption and emission \citep{2019ApJ...882L..23D,2019ApJ...887..257D}. This has been followed by detections in emission over more sightlines spread across the sky \citep{2022arXiv220109915G,2022arXiv220802477B}. As the evidence in favor of the hot component mounts, it is imperative to understand its elemental makeup, spatial distribution and its relationship with the other components of the CGM.

In \citet{2021ApJ...918...83D}, hereafter referred to as D21, the warm-hot and hot components of the CGM of the Milky-Way have been detected in the sight line of \mrk{} in absorption. While the warm-hot phase had already been detected in both emission and absorption \citep{2007ApJ...658.1088Y,2014PASJ...66...83S,2014Ap&SS.352..775G} towards this sight line, this marks the first time the hot phase has been found in an antigalactic sightline. More so, deviations from a solar chemical composition was noticed. This necessitates a follow-up, studying the soft X-ray emission of components of the CGM. 

While \mrk{} being a nearby bright blazar ensures that it has been one of the most well-studied blazars with abundant archival observations, there exists a serious drawback due to the fact that it also very bright. In \S \ref{sec:data} we explore our dataset and the methods undertaken to remove the central emission from \mrk{} that can otherwise bias the spectral signatures of the faint emission from the CGM. Following this, the spectra is fit with emission models that take into account various foreground and background sources of soft X-ray emission, as detailed in \S \ref{sec:xspec}.

In \S \ref{sec:results} we discuss the results of the spectral-fitting procedure and characterize the CGM using the temperature and emission measure (EM) from the best-fit spectral model. We discuss the implications of these results on our understanding of the CGM in \S \ref{sec:discussion} and detail the conclusion in \S \ref{sec:conclusion}. Hereafter, in this paper we use the terms ``warm-hot'' and ``virial temperature'' interchangeably to describe the $\sim 10^6$ K gas while the terms ``hot'' and ``super-virial temperature'' are used to imply the $\sim 10^7$ K gas.

\section{Datasets} \label{sec:data}

\subsection{Blazar field}

\begin{figure*}
\centering
\includegraphics[scale=0.275,clip]{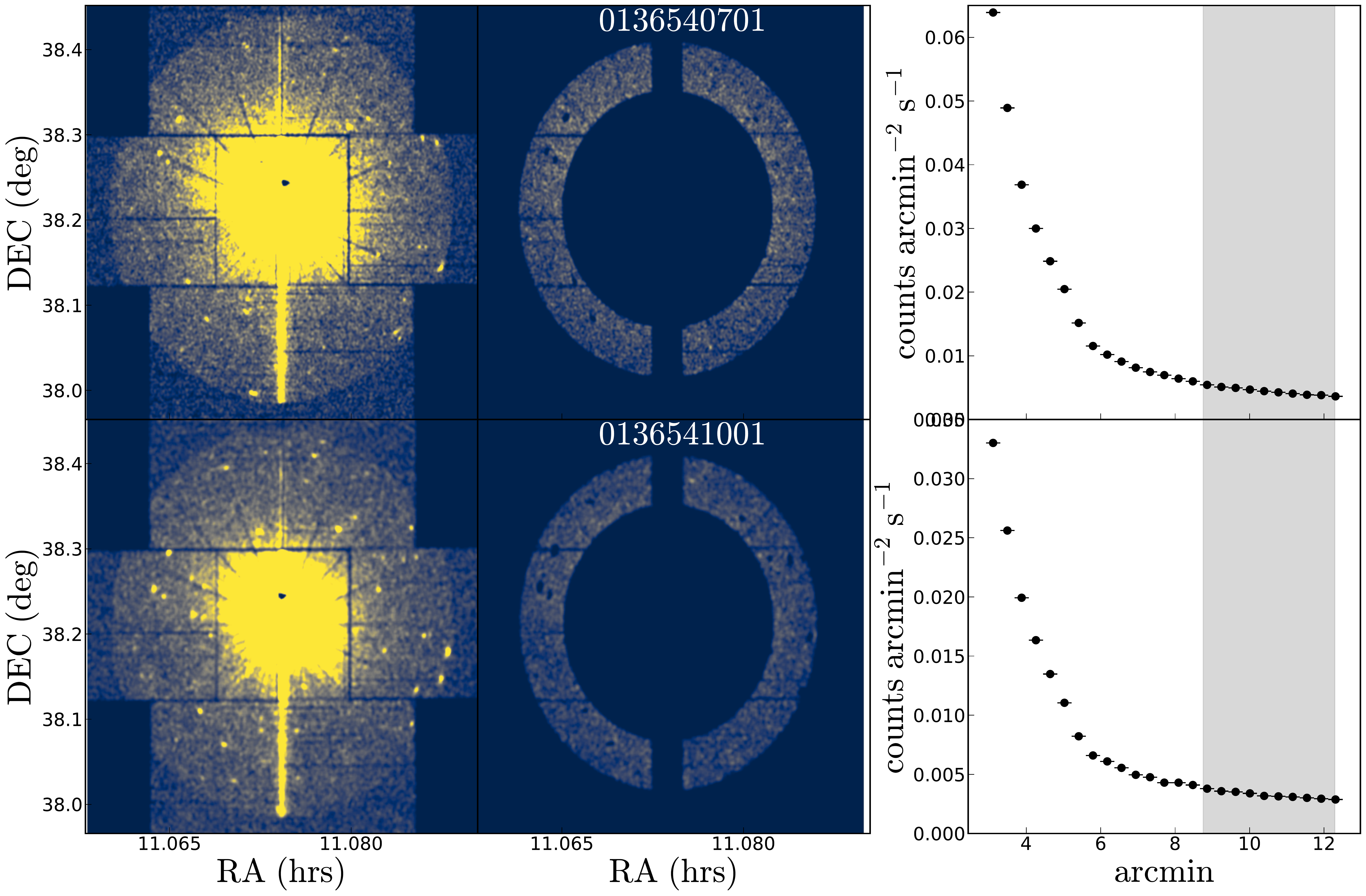}
\caption{\textit{Left}: Raw EPIC-MOS2 images for the two exposures with obsids 0136540701 (\textit{upper row}) and 0136541001 (\textit{lower row}) showing the stray light from \mrk{} at the center and other point sources in the background. \textit{Center}: The disjoint annular region in the photon energy band 0.35-7.0 keV with point sources masked and with minimal contamination from the blazar, from which the spectrum is extracted. \textit{Right}: The radial flux density profiles is shown using the \textit{black} points with the \textit{gray} shaded region representing the extent of the annulus where contamination of the blazar is $< 2\%$ of the total flux density in the FoV. The vertical errors represent the the Poisson uncertainty of the flux density with the horizontal error bars representing the bin-widths. }
\label{Figure:fov_im_profile}
\end{figure*}

We searched the XMM-Newton Science Archive for EPIC observations with \mrk{} in its field of view (FoV), considering only those exposures which are in the Full Frame imaging mode. This mode involves all the CCDs of the EPIC-pn camera being activated, leading to the full accessible FoV of 30'x30' that is ideal for observing the diffuse CGM after excluding the emission from the central source. The two cameras aboard XMM-Newton- EPIC-pn and EPIC-MOS2 have unique advantages for each. While EPIC-pn has a superior collecting area, EPIC-MOS has a superior spectral resolution. We choose MOS2 data for this study because there are no Full Frame pn observations with exposure time of  $>20$ks.
 There are 2 Full-Frame exposures with obsid:0136540701 and obsid:0136541001 from MOS2 totalling $\sim 140$ks in duration.  

The details of these exposures are tabulated in Table \ref{Table:obs}. We use the XMM-Extended Source Analysis Software (ESAS) \citep{ESASCookbook} to reduce the data. We use the $\texttt{mos-filter}$ to filter out the out-of-time (OoT) events, use $\texttt{cheese}$ routine to identify and excise the point sources, and use $\texttt{mos-spec}$ and $\texttt{mos\_back}$ for extraction of the spectra and detector background respectively. We follow exactly the same procedure as followed in \citet{2019ApJ...887..257D} except for the way the contamination from \mrk{}, is removed. This is because, relative to 1ES 1553+113, Mrk 421 is a closer and brighter source \citep{2005A&A...433.1163D} and is also variable \citep{2020ApJ...897...25B}, hence we attempt to rigorously  eliminate any contamination from Mrk 421 on its surrounding field.  

The raw image around \mrk{} is depicted in the left panel of Figure \ref{Figure:fov_im_profile}, where photons from \mrk{} saturate pixels not only at the center but even in those beyond the $12.5^{\prime \prime}$ FWHM of the EPIC-pn camera. Therefore to minimize stray light contamination from \mrk{} we proceed by computing the flux density profile as a function of angular separation from the center of the blazar, as seen in the right pane of Figure \ref{Figure:fov_im_profile}. We observe that beyond $9^{\prime}$ the contamination from \mrk{} is negligible.  To exactly determine the excision radius we subtract the density at the outermost radius, where the profile has flattened out, from the whole profile to give the excess flux density that is attributed to the blazar's emission. The uncertainty on the density estimate is calculated assuming the flux follows a Poisson distribution and the uncertainty on the excess flux is approriately propagated. The excision radius is chosen to be $8.75^{\prime}$, beyond which the contamination from the blazar, i.e., the excess flux density is $\sim 2\%$ the total flux density in the FoV. The gray shaded region in Figure \ref{Figure:fov_im_profile} represents the region beyond the excision radius that we use in our analysis.

We also exclude the region beyond 12.29', to account for vignetting, i.e., decreased response near the edges and also for greater soft proton contamination. There also exists readout streaks, oriented vertically in the detector coordinates and these are excluded using rectangular boxes. We are left with a disjoint annular region within 8.75' and 12.29', seen in the central pane of Figure \ref{Figure:fov_im_profile} from where we extract the spectra.

\begin{table}
\centering
  \caption{Details of the EPIC-MOS2 observations centered on \mrk{}.  \label{Table:obs}}
  \begin{tabular}{|c|cc|} \hline
ObsID &Good Time &Signal-to- \\
&Interval &Noise,\\
	&(ks) &$S/N$ \\
\hline
0136540701 &48.31 &159.66 \\
0136541001 &56.94 &120.19 \\
\hline 
\end{tabular}  
\end{table}


\subsection{Auxiliary fields}

We searched the XMM-Newton Science Archive for observations that were beyond the immediate vicinity of \mrk{}, but within five degrees of the blazar. Doing so we can further test the presence of the hot component of the CGM independent from any contamination from the blazar. Furthermore, we can study the spatial variations of the characteristics of this component at small angular scales on the sky. We found good quality data of distant clusters and quasars with $S/N > 200$ .

The dataset from EPIC-pn thus obtained in provided in Table \ref{Table:PNenvironlist}. The emission from the point sources in the FoV for each of the observations are effectively removed by the application of the $\texttt{cheese}$ routine. The targets (clusters/AGNs) were masked out using circle or ellipse shaped regions in $DS9$. We manually inspect the images for any additional diffuse emission from the intercluster medium and find that this resulted in little contamination over the FoV. This was followed by the $\texttt{pn-filter}$ and $\texttt{pn\_back}$ command for extraction of the spectra and its background respectively.

\begin{table*}
\centering
  \caption{Details of the EPIC-pn exposures in the auxiliary fields located within 5 deg from \mrk{}. \label{Table:PNenvironlist}}
  \begin{tabular}{|cc|cc|cc|cc|} \hline 

ObsID &Target &Galactic &Galactic &Angular sep.  &Abs. column 	&Good Time &Signal-to- \\
&Name &lat., $l$ &long., $b$ &from Mrk421, &density, $N_H$   	&Interval &Noise,\\
& &(degrees) &(degrees) &$D$ (degrees) &(10$^{20}$ cm$^{-2}$) 	&(ks) &$S/N$ \\
\hline
0694651001 &ASASSN-14ae &189.09 &66.90 &4.20 &1.61  &15.84 &245.56 \\
0694651101 &  &  & & &   &19.60 &286.00 \\
0694651601 &   &  & &  & &12.70 &226.03 \\
\hline
0205370201 	&CL1115.8+4239       &168.02 &65.15 &4.97 &3.50  &21.69 &210.21 \\
\hline
0205370101 &CL1103.6+3555  &185.13 &65.51 &2.27 &2.70 &26.65 &231.91 \\
\hline
0842730201 &PGC032873 &172.45 &62.09 &4.41 &1.19    &60.36 &501.65 \\
0842730101 &  &  & & &  &59.80 &503.09 \\
\hline
0841481701 &2MASXJ10514428       &186.93 &63.19 &3.60 &2.20 &77.55 &597.51 \\
\hline 
\end{tabular}  
\end{table*}

\section{Spectral Modeling \& Fitting} \label{sec:xspec}

We removed the emission from \mrk{} and other bright point sources in the FoV, and extracted the spectra from regions shown in Figure \ref{Figure:fov_im_profile} for each of the two observations. The spectra from these ``empty'' fields contain contributions from instrumental lines, unresolved compact sources in the background and foreground sources, such as the Local bubble, in addition to the emission from the MW CGM itself. Identifying and quantifying the spectral signatures of these contaminants is required for a robust study of the hot gaseous halo of the Milky-Way.

We use the $\texttt{grppha}$ routine to rebin the raw spectra obtained such that each energy bin has $50$ data points, thereby increasing the $S/N$ per bin. While fitting the model spectra we only consider the 0.35 keV to 7.0 keV bandwidth and ignore all bins with ``bad" data. 


\subsection{Instrumental Lines}

The instrumental lines are the result of fluorescent emission from instruments around the detector or the detector itself. For example, the spectra from the MOS2 detector feature two prominent lines- the Al K$\alpha$ at 1.48 keV and the Si K$\alpha$ line at 1.74 keV. We model both of these lines as unabsorbed zero-width Gaussians \citep{2019ApJ...887..257D}, allowing their normalizations to vary. If there are multiple observations of the same object in a single epoch, for each of them we tie the normalization of each line to a single value. 

The pn spectra are expected to have the Al K$\alpha$ line only, but we found that modelling the spectrum between 1.47-1.5 keV with a single line feature was insufficient. Thereby we added a second Gaussian at 1.49 keV for the unidentified emission and this was found the improve the quality of fit significantly. 

\subsection{Soft Proton Background}

The soft proton background emanates from the high-energy protons trapped in Earth's magnetic field  \citep{2018ApJ...867....9F} interacting with the detector. We model this contribution as an unabsorbed broken power-law with a break at 3 keV \citep{ESASCookbook}. The response matrix of the appropriate EPIC detector is not folded with this model. The power-law indices and normalizations are allowed to vary for both the observations.

\subsection{Galactic Absorption}

The galactic absorption is modeled using the $\texttt{phabs}$ model in XSPEC with the absorbing column density N(HI) along a sight line queried using the $nH$ \href{https://heasarc.gsfc.nasa.gov/cgi-bin/Tools/w3nh/w3nh.pl}{interface} \footnote{\href{https://heasarc.gsfc.nasa.gov/cgi-bin/Tools/w3nh/w3nh.pl}{https://heasarc.gsfc.nasa.gov/cgi-bin/Tools/w3nh/w3nh.pl}}
on HEASARC \citep{1990ARA&A..28..215D,2005A&A...440..775K,2016A&A...594A.116H}. While in the sight line of \mrk{} this is fixed at $1.34\times 10^{20}$ cm$^{-2}$, for the auxillary observations the respective values listed in Table \ref{Table:PNenvironlist} is used. The $\texttt{phabs}$ model is convolved with the non-instrumental and non-foreground components of the model spectrum.

\subsection{Foreground}

The spectrum of the diffuse emission is dominated by foreground and background. Following \citet{2012ApJ...756L...8G,2022arXiv220109915G} we model the  foreground arising from the Local Hot Bubble (LHB). We assume that the emission of the LHB arises from a collisionally-ionized plasma at a single temperature $T$ with solar abundances prescribed by \citet{2009ARA&A..47..481A} using the APEC model \citep{2001ApJ...556L..91S}. We follow the measurements of \citet{2017ApJ...834...33L} and keep the temperature of the LHB  fixed at 0.097 keV whereas the normalization is kept free.

\subsection{Cosmic X-ray Background}

The cosmic X-ray background \citep{1962PhRvL...9..439G} comprises X-ray light mainly from the unresolved AGNs in the universe \citep{2007A&A...463...79G}. This is modeled as a power law with absorption from the Galactic ISM \citep{2010ApJ...723..935H} with the normalization and the power-law index left as free parameters.

\subsection{Galactic Halo emission}

The soft X-ray emission from the Galactic halo gas, in other words, the CGM of the Milky-Way is the signal of our interest. This is modeled as an absorbed collisionally ionized plasma in thermal equilibrium using the APEC model. The assumption of collisional ionization equilibrium (CIE) is justified given that simulations both cosmological as well as of individual galaxies \citep{2010MNRAS.408.2051D,2012MNRAS.425.1270S,2010ApJ...724.1044S}. Previous studies \citep{2010ApJ...723..935H,2012ApJ...756L...8G,2018MNRAS.474..696G} validate this assumption.

\subsection{Field of View area}

The emission measure (EM) is calculated from the best-fit APEC normalization $N_{APEC}$ \footnote{The norm $N_{APEC}$ provided in the XSPEC \href{https://heasarc.gsfc.nasa.gov/xanadu/xspec/manual/XSmodelApec.html}{manual} is inverted and appropriate units are filled in.} by applying the following formula:

\begin{equation}
\left( \frac{EM}{cm^{-6} \ pc} \right) = \left( \frac{N_{APEC}}{1961 \ cm^{-5}} \right) \left( \frac{\Omega}{sr} \right)^{-1},
\end{equation}

where $\Omega$ is the solid-angle of the region from which the spectra is extracted. In this work, these regions have complicated geometries, a disjoint annulus for the MOS2 observations and with point sources masked with $\texttt{cheese}$ in both the MOS2 and pn exposures. Therefore, we calculate the on-sky projected areas of these regions by counting all pixels in the region which detect at least one photon. To get the area subtended by this region on the sky in units of steradians, we scale the number by the area subtended by each pixel which is 1.469$\times10^{-10} sr$ \citep{ESASCookbook}.

To validate the accuracy of this method, we attempt to compare its estimate with a calculation of the disjoint annulus using simple geometric considerations. For exposures 0136540701 and 0136541001, the pixel based estimate was found to be 92.45\% and 91.89\% of the geometry based estimate respectively. This mismatch mainly corresponds to the effect of the masked point sources. Therefore, when calculating the uncertainties on the EM estimate, we propagate errors from the normalization uncertainty. In light of the comparison, we assume that the fractional uncertainty on $\Omega$ is 10\%:

\begin{equation}
\frac{\sigma_{EM}}{EM} = \sqrt{\left( \frac{\sigma_{N_{apec}}}{N_{apec}} \right)^2 + (0.1)^2}.
\end{equation}

\subsection{Fitting}

All the spectral fitting was done using XSPEC \footnote{\href{https://heasarc.gsfc.nasa.gov/xanadu/xspec/}{https://heasarc.gsfc.nasa.gov/xanadu/xspec/}}. For each field we fitted the spectra from different exposures simultaneously to obtain better constraints on the fitted parameters.  The normalization of the Gaussian lines was seen to vary with time. Similarly, the normalization of the LHB component is known to be anisotropic, although weakly. For these reasons, we kept the normalization of the Gaussian lines and foreground components separate for exposures from different epochs or different fields.
Due to the  different role angles, different exposures around the same target covered slightly different fields (for example, see left pane of Figure \ref{Figure:fov_im_profile}). This resulted in differences in solid angle $\Omega$ used to extract the CGM spectrum; we weighed the models to account for this difference. We chose the solid-angles normalized w.r.t. their mean, i.e., $ \Omega_{i}/ \langle \Omega \rangle$ as the weight for the $i$th exposure; this noticeably improved the fit quality compared to that without using a weight.


We began the spectral fits with a single temperature or 1-T model of the Galactic halo emission.  We then freeze all the parameters of the best-fit 1-T model except the normalizations of the different foreground and background components and the temperature and the normalization of the APEC model. Following this we added a second component of APEC, also absorbed by the Galactic ISM, to our model, effectively creating a two temperature or 2-T model for the CGM. We sequentially introduce and fit the two components of the CGM instead of doing so at once due to two considerations. First, the EM of the warm-hot component is expected to be larger than the hot component \citep{2019ApJ...887..257D}. Second, the quality of the data does not allow to consistently fit the two components in the spectrum at once. Our method allows us to obtain better constraints of the model parameters, and we test for the statistical significance of adding an extra component, as discussed further below. 


\section{Results} \label{sec:results}

\subsection{Blazar field}

\begin{figure*}
\centering
\includegraphics[scale=0.45,clip]{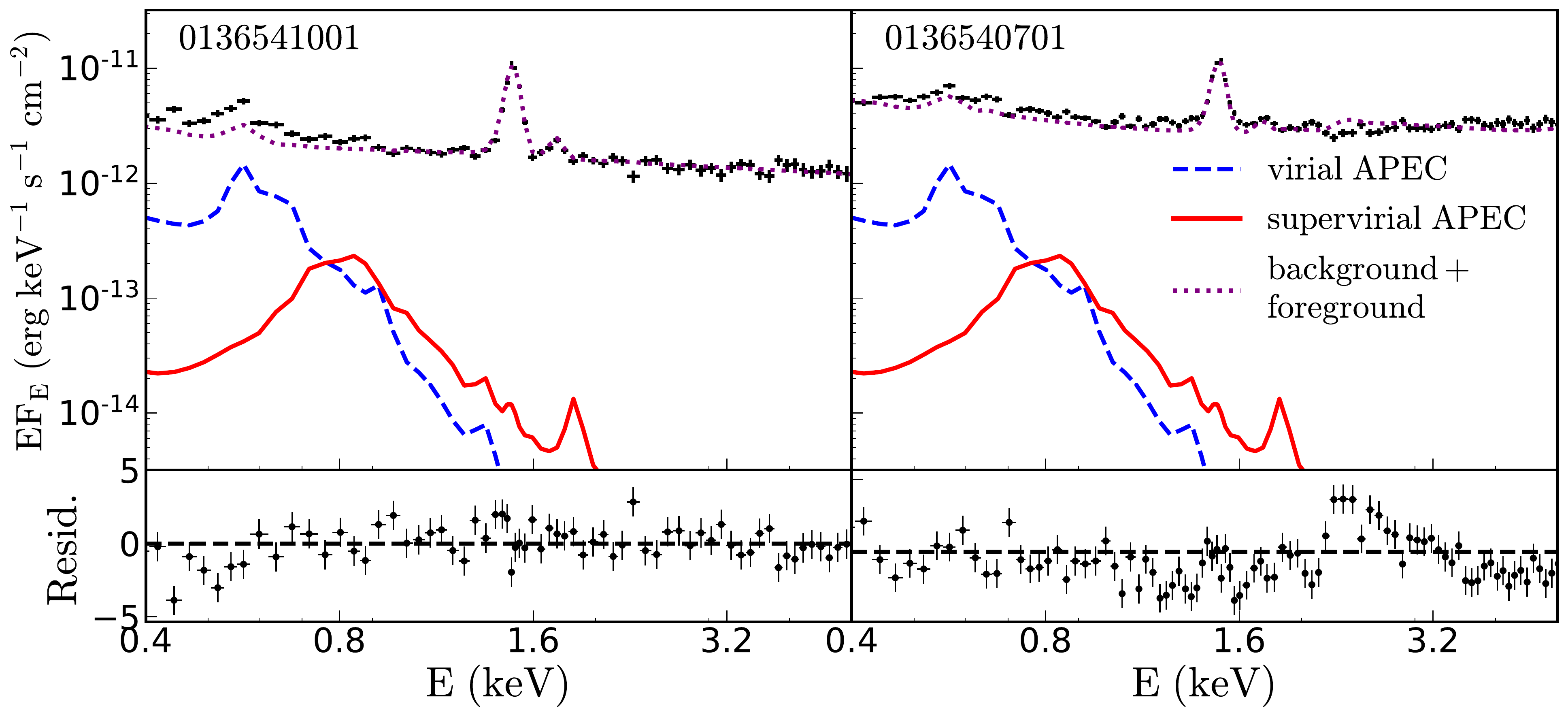}
\caption{The spectra obtained from exposures 0136541001 (\textit{left}) and 0136540701 (\textit{right}) in the energy band 0.4-5.0 keV are shown using the \textit{black points} with errorbars. The \textit{dashed blue }lines and the \textit{solid red} lines represent the APEC models corresponding to the virial and super-virial phases with log ($T/$K) = $6.33_{-0.02}^{+0.03}$ and log ($T/$K) = $6.88_{-0.07}^{+0.08}$ respectively. The total contribution from the rest of the background and foreground components is plotted as the \textit{dotted purple} line.}
\label{Figure:mos2_spec}
\end{figure*}

\begin{table*}
\centering
  \caption{Characteristics of the CGM resulting from fitting the EPIC-MOS2 spectra from the blazar field \label{Table:mos2_scatter1}}
  \begin{tabular}{|ccc|ccccc|c|} \hline 
  
\multicolumn{3}{|c|}{One temperature CGM model} &\multicolumn{5}{c|}{Two temperature CGM model} & \\
\hline

$kT_1$ &EM$_1\times 10^3$ &$\chi^2$ &$kT_1$ &EM$_1\times 10^3$	&$kT_2$ &EM$_2\times 10^3$  &$\chi^2$ &F-test  \\
(keV) &(cm$^{-6}$ pc) &$/d.o.f.$ &(keV) &(cm$^{-6}$ pc)	&(keV) &(cm$^{-6}$ pc) &$/d.o.f.$ &prob. (\%)  \\
\hline
 & & & & & & & &  \\

$0.202\pm0.009$ &$3.75_{-0.50}^{+0.49}$ &$1161.42$ &$0.186_{-0.010}^{+0.017}$ &$4.01_{-0.55}^{+0.54}$ &$0.66_{-0.10}^{+0.12}$ &$0.45_{-0.17}^{+0.15}$ &$1157.04$ &$82.14$ \\

 & &$/902$ & & & & &$/900$ & \\
\hline



\end{tabular}  

\end{table*}


The best-fit parameters that we obtained for the Galactic halo emission have been tabulated in Table \ref{Table:mos2_scatter1}. The uncertainties on the best-fit temperature and normalization correspond to the 1$\sigma$ confidence intervals that is calculated by freezing all of the other model parameters.

The best-fit 1-T model gives us $kT = 0.20 \pm 0.01$ keV and EM=$3.75_{-0.50}^{+0.49}$\emu. For reference we have the estimates from \citet{2014Ap&SS.352..775G} made by applying the same model on two sight lines $\sim 1.4$ deg away from \mrk{} which are, $kT = 0.180 \pm 0.012$ keV and EM$= (2.5 \pm 0.6)$\emu. Considering these we find that both the parameters are non-overlapping within their $1\sigma$ uncertainties. Moving to a 2-T model we find $kT = 0.19_{-0.01}^{+0.02}$ keV and EM=$4.01_{-0.55}^{+0.54}$\emu{} for the virial phase and $kT = 0.66_{-0.10}^{+0.12}$ keV and EM=$0.45_{-0.17}^{+0.15}$\emu{} for the super-virial phase, with a improvement in the $\chi^2/$d.o.f..  The F-test is applied to discern whether a 2-T model is favored over the 1-T model, thereby assessing whether the super-virial phase is required. The test returns a probability of $82.14 \%$ favoring the 2-T model.



Studying the absorption around \mrk{} sightine, \citetalias{2021ApJ...918...83D} arrived at estimates of $kT = 0.129 \pm 0.009$ keV and $kT = 2.76^{+1.29}_{-0.43}$ keV for the virial and super-virial components respectively. These temperatures bracket our estimates using the emission based study, but they do not overlap. We discuss this further in Sections \ref{sec:emiss_bias} and \ref{sec:hot_how}.

\subsection{Auxiliary fields}

\begin{figure}[h!]
\centering
\includegraphics[scale=0.2,clip]{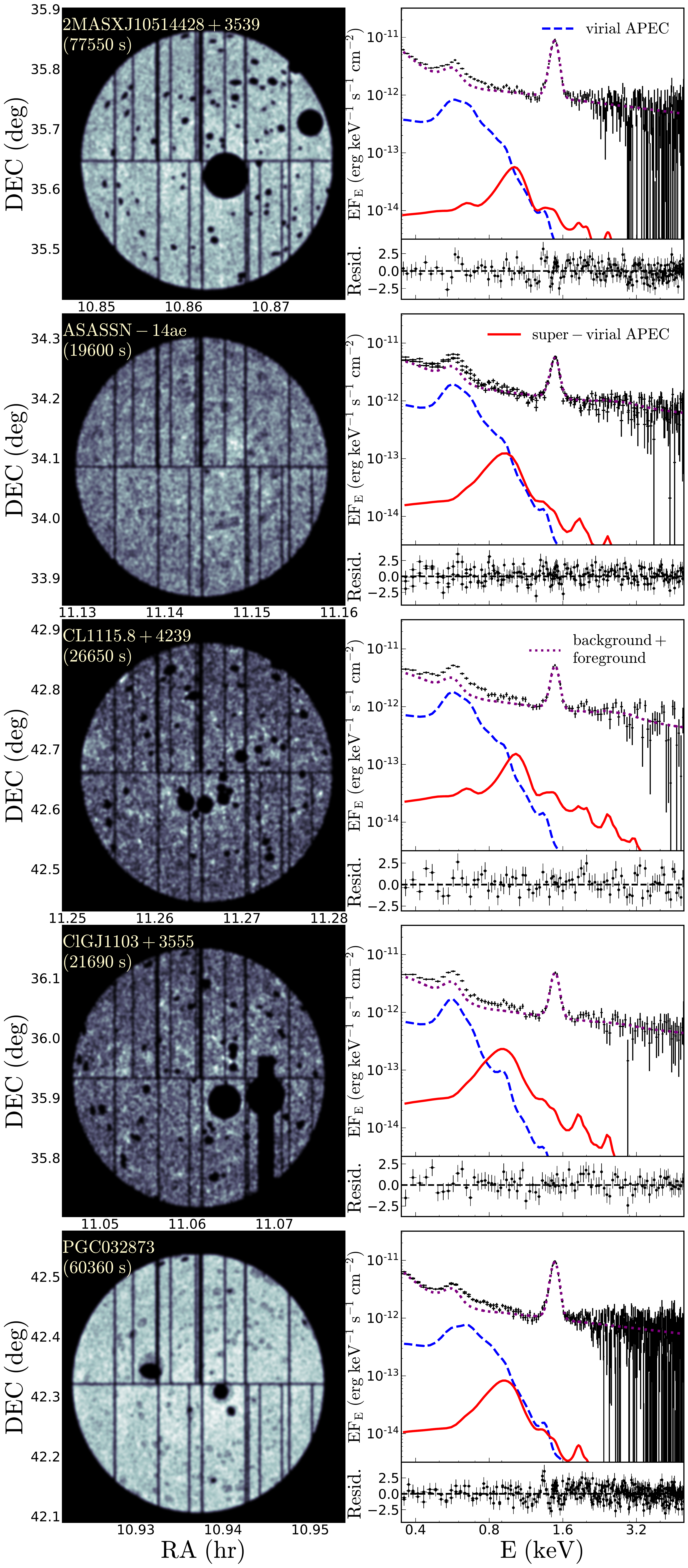}
\caption{Images of the masked FoV (\textit{left}), from which the spectra (\textit{right}) are extracted. The \textit{dashed blue} lines and the \textit{solid red} lines represent the APEC models corresponding to the virial and super-virial phases detected along each sight lines. The total contribution from the rest of the components are plotted as the \textit{dotted purple} lines.}
\label{Figure:esd_mock}
\end{figure}

\begin{table*}
\centering
  \caption{Characteristics of the MW CGM resulting from fitting the EPIC-pn spectra from the auxiliary fields \label{Table:pn_environ_scatter}}
  \begin{tabular}{@{}|l|ccc|ccccc|c|} \hline 
  
 &\multicolumn{3}{c|}{One temperature CGM model} &\multicolumn{5}{c|}{Two temperature CGM model} & \\
\hline
  
 Field &$kT_1$ &EM$\times 10^3$	&$\chi^2$ &$kT_1$ &EM$\times 10^3$	&$kT_2$ &EM$\times 10^3$  &$\chi^2$ &F-test \\
&(keV) &(cm$^{-6}$ pc) &$/d.o.f.$ &(keV) &(cm$^{-6}$ pc) &(keV) &(cm$^{-6}$ pc) &$/d.o.f.$ &prob. (\%) \\
 \hline
& & & & & & & & & \\

All &$0.221\pm0.003$ &$1.42\pm0.15$ &$6766.748$ &$0.219_{-0.003}^{+0.004}$ &$1.52\pm0.16$ &$1.03_{-0.07}^{+0.09}$ &$0.14\pm0.03$ &$6755.148$ &$>99.99$ \\
& & &$/6538$ & & & & &$/6540$ & \\
\hline
\hline
ASASSN &$0.192\pm0.004$ &$ 2.54 \pm 0.27$ &$1163.159$ &$0.191_{-0.006}^{+0.005}$  &$2.89 \pm 0.31$  &$0.93 \pm 0.19$ &$0.14\pm0.05$ &$1160.025$ &$92.20$\\ 
-14ae & & &$/1153$ & & & & &$/1152$ & \\
CL1103.6 &$0.155_{-0.009}^{+0.008}$ &$2.94_{-0.43}^{+0.50}$ &$454.399$ &$0.159\pm0.009$ &$3.28_{-0.44}^{+0.50}$  &$0.89_{-0.10}^{+0.11}$ &$0.28\pm0.07$ &$449.04$ &$98.17$ \\
+3555 & & &$/471$ & & & & &$/470$ & \\
CL1115.8 &$0.190 \pm 0.005$ &$3.00 \pm 0.35$ &$446.611$ &$0.193\pm0.005$ &$3.12_{-0.35}^{+0.36}$ &$1.30_{-0.21}^{+0.46}$ &$0.33_{-0.15}^{+0.27}$ &$444.429$ &$85.67$  \\
+4239 & & &$/439$ & & & & &$/438$ & \\
PGC03- &$0.245\pm0.006$ &$1.22 \pm 0.13$ &$1892.661$ &$0.239_{-0.009}^{+0.008}$  &$1.22 \pm 0.13$ &$0.92_{-0.15}^{+0.18}$ &$0.10 \pm 0.04$  &$1888.363$ &$95.98$ \\ 
2873 & & &$/1853$ & & & & &$/1852$ & \\
2MASXJ- &$0.212\pm0.006$ &$1.44 \pm 0.16$ &$988.211$  &$0.210\pm0.007$ &$1.42 \pm 0.16$ &$1.21_{-0.27}^{+0.44}$ &$0.11_{-0.07}^{+0.10}$ &$985.309$ &$90.03$ \\
10514428 & & &$/923$ & & & & &$/922$ & \\

\hline 
\end{tabular}  
\end{table*}

The results of spectral fitting of the data from the 5 auxiliary fields within five degrees  of \mrk{} are listed in Table \ref{Table:pn_environ_scatter}. We fit all the five spectra from the auxiliary  fields simultaneously, first with a 1-T model, followed by a 2-T model. 
The best-fit parameters of the 2-T model are $kT_1 = 0.219_{-0.003}^{+0.004}$ keV and EM $= 1.52\pm0.16$ \emu for the virial component, and $kT_2 = 1.03_{-0.07}^{+0.09}$ keV and EM $= 0.14\pm0.03$ \emu{} for the super-virial component.  The inclusion of the super-virial phase is strongly favored by a F-test. 


Encouraged by these results, we fitted the spectra of each of the five fields separately. The best-fit temperature and EM measurements of the individual sight lines are similar to each other and to what we got from the blazar field, albeit a few things stand out. The virial phase has a range of temperatures $\rm 0.16 keV < kT_1 < 0.24 keV$ and the super-virial phase range is $\rm 0.89 keV < kT_1 < 1.30 keV$.  The F-test probabilities for the inclusion of the second phase are all $>85.67$\%. All the auxiliary fields have lower EM for the virial component and a higher temperature of the super-virial component compared to the blazar field (see Figure \ref{Figure:all_scatter}). 

The best-fits from the fields around PGC032873 and 2MASXJ10514428 yield an EM of the virial component that is $\approx 3\times$ lower compared to the other fields. Since these are high $S/N$ observations with long exposures, they drive down the virial component EM in the simulteneous fit as well. The scatter of these estimates in the parameter space can be either due to some unaddressed systematic or an actual variation of the EM of the virial temperature gas across the sky at angular scales $<10$ deg. In Section \ref{sec:patchy} we discuss the causes and implications of the anisotropy of the EM and temperature of the MW CGM.


\subsection{All fields}

\begin{figure*}[h!]
\centering
\includegraphics[scale=0.3,clip]{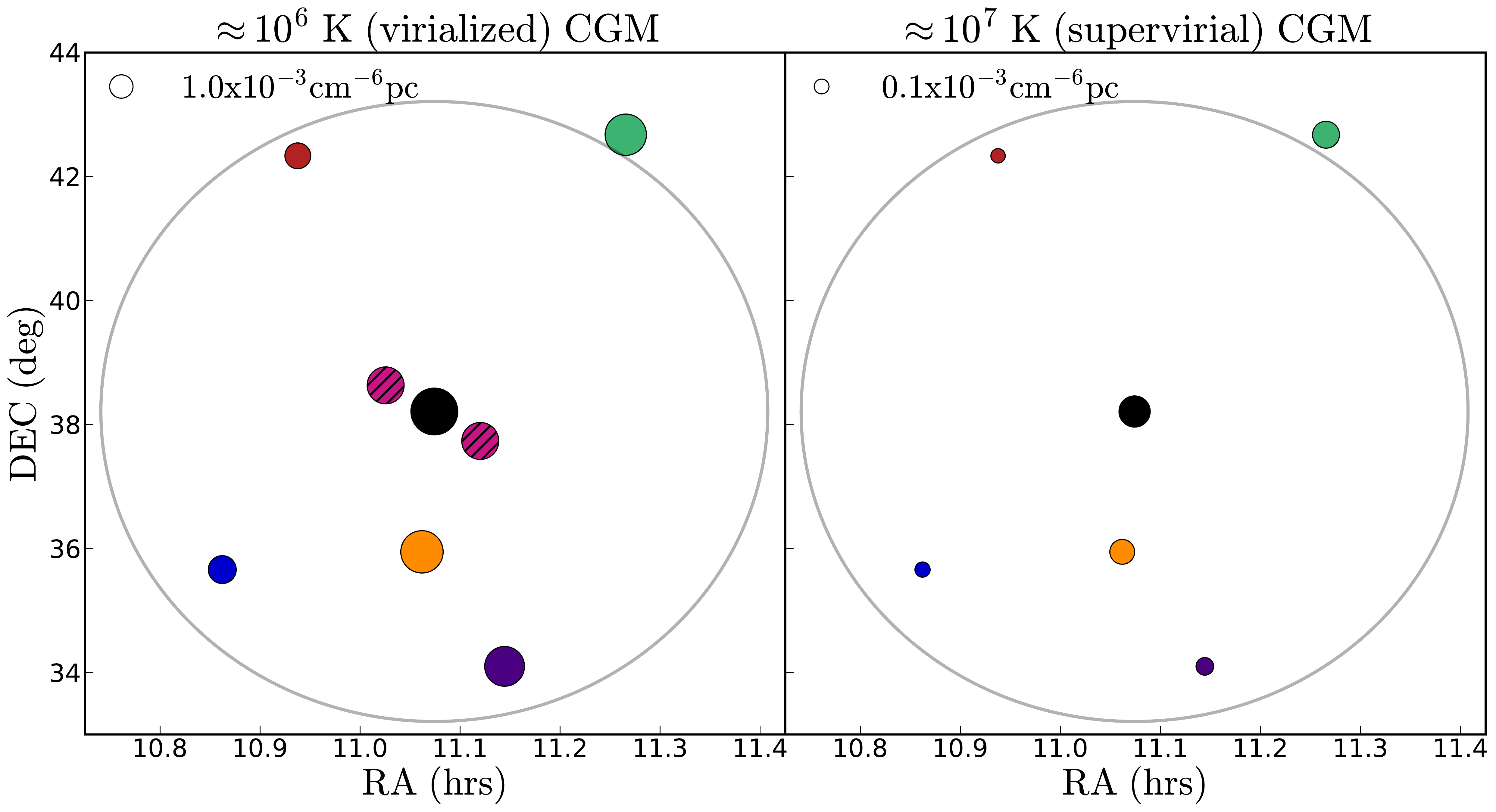}
\caption{On-sky maps of the auxiliary fields located around the \mrk{} sightline, with the \textit{left} and \textit{right} panels showing the virial and super-virial components of the CGM respectively. The sizes of the points are scaled by the EM. The grey circle is of radius 5 deg and centered on \mrk{}. The measurements of the virial component taken from \citet{2014Ap&SS.352..775G} have been shown using the the two hatched points for reference.}
\label{Figure:aux_param_map}
\end{figure*}

\begin{figure*}
\centering
\includegraphics[scale=0.4,clip]{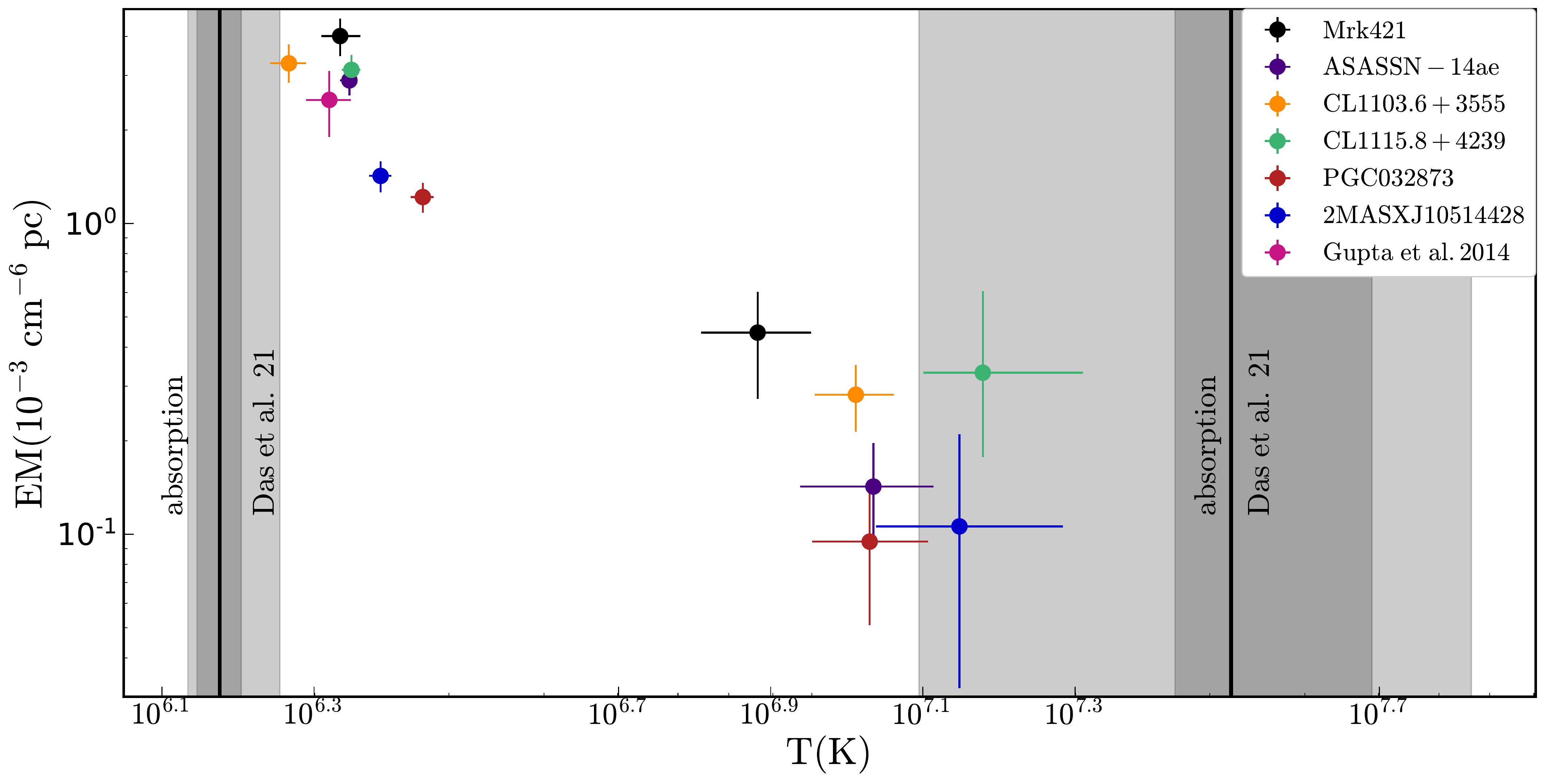}
\caption{The best-fit temperature and EM of the MW CGM in all the fields. These include the primary field around \mrk{} (\textit{circle}), the 5 auxiliary fields labeled according to the target object, as well the results from \citet{2014Ap&SS.352..775G}. The absorption based temperature estimates and the associated 90\% and 99\% confidence intervals from \citetalias{2021ApJ...918...83D} are marked as the vertical solid \textit{black} lines and the shaded \textit{gray} regions respectively. \label{Figure:all_scatter}}
\end{figure*}

The temperature and EM estimates we found from spectral fits along the different sight lines are shown as points in Figure \ref{Figure:all_scatter}. The plot includes the results for the field around \mrk{} using the EPIC-MOS2 spectra, and the 5 auxiliary fields with spectra from the EPIC-pn camera. Furthermore, the plot shows the measurements from \citet{2014Ap&SS.352..775G} from fields close to the \mrk{} sightline and temperatures based on absorption  along the \mrk{} sightline from \citetalias{2021ApJ...918...83D}. There are two clusters of points at low and high temperatures which belong to the virial and super-virial components of the CGM respectively. We observe that there is a considerable scatter in the EM estimates of the virial component. For both of these components, there is scatter among the fields, which we discuss in detail later. Furthermore, we spatially visualize the characteristics of the two components of the CGM in Figure \ref{Figure:aux_param_map} where the the strength of the emission in each field has been depicted by the size of the circles. This again shows that there is a scatter in the EMs of different fields for both virial and super-virial phases. Note also that the EM does not decrease monotonically away from the blazar field; once again this confirms that the contamination from the blazar, if any, is negligible. 

\subsection{Neon Over-abundance} \label{sec:abundances}

The Ne IX K$\alpha$ line is at 0.905 keV and the Ne X Ly$\alpha$ line is at 1.021 keV \citep{2012ApJ...756..128F}. If Ne is over-abundant in a virial temperature gas with $kT \approx 0.2$ keV, it can mimic the spectral signature of a super-virial temperature gas with $kT \approx 1.0$ keV. The Ne IX K$\alpha$ line is more prominent, and beyond 1.0 keV the CXB power-law drowns  the Ne X Ly$\alpha$ line. In \citet{2009ApJ...690..143Y}, the diffuse soft X-ray emission along multiple sight lines are fit with variable Ne and Fe abundances. Models with $\approx 3\times $ the solar Ne abundance are favoured, and these are found to be degenerate with models comprising a hotter gas with $0.6<kT<0.9$ keV. This is also in agreement with \citet{2016ApJ...816...33U}; the spectra from the NPS region was found to have excess soft X-ray emission around 0.9 keV that could be either explained by a $kT = 0.29$ keV gas with Neon abundance of Ne/O=1.7 or a hotter gas with $kT =\approx 0.76$ keV. \citet{2021ApJ...909..164G} also identify the emission from 0.8-1.0 keV as either stemming from a Neon enhanced virial temperature gas with 1.5<Ne/O<4.0 or a 2-T model with a solar abundances and a super-virial temperature gas at 0.65$<kT<$0.90 keV.

We attempted to test the hypothesis of Neon overabundance by first fitting 1-T vAPEC models to all the spectra. Fitting  each field separately, we obtained 1.1< Ne/O <4.8. All of the fields favored the 2-T model with F-test probabilities of >98\%. Fitting all the spectra simultaneously, we find that Ne/O = 1.98 could fit the data, but it was less favorable compared to the 2-T model with solar abundances, as validated by the F-test probability favoring the latter by 99.93\%. Thus in this work we resolve the degeneracy between the two models and conclude that a 2-T model is required. 


\subsection{Unresolved Nitrogen lines} 

\citet{2019ApJ...882L..23D} found an excess around 0.515 keV in the MOS2 spectra compared to the pn spectra of the same field. This could arise from pn not being able to spectrally resolve the N VII Ly$\alpha$ and/or N VI K$\beta$ lines. Indeed, adding these lines to the model improved the fit. We therefore tried variable nitrogen abundance in our models as well. Fitting the MOS2 spectra simultaneously with a 2-T model, we got N/O=4.27930 with a 98.11\% F-test probability. This indicates that the virial phase has a super-solar Nitrogen abundance. 

\subsection{Effect of Fe/O on the measurements}

\begin{figure}
\centering
\includegraphics[scale=0.375,clip]{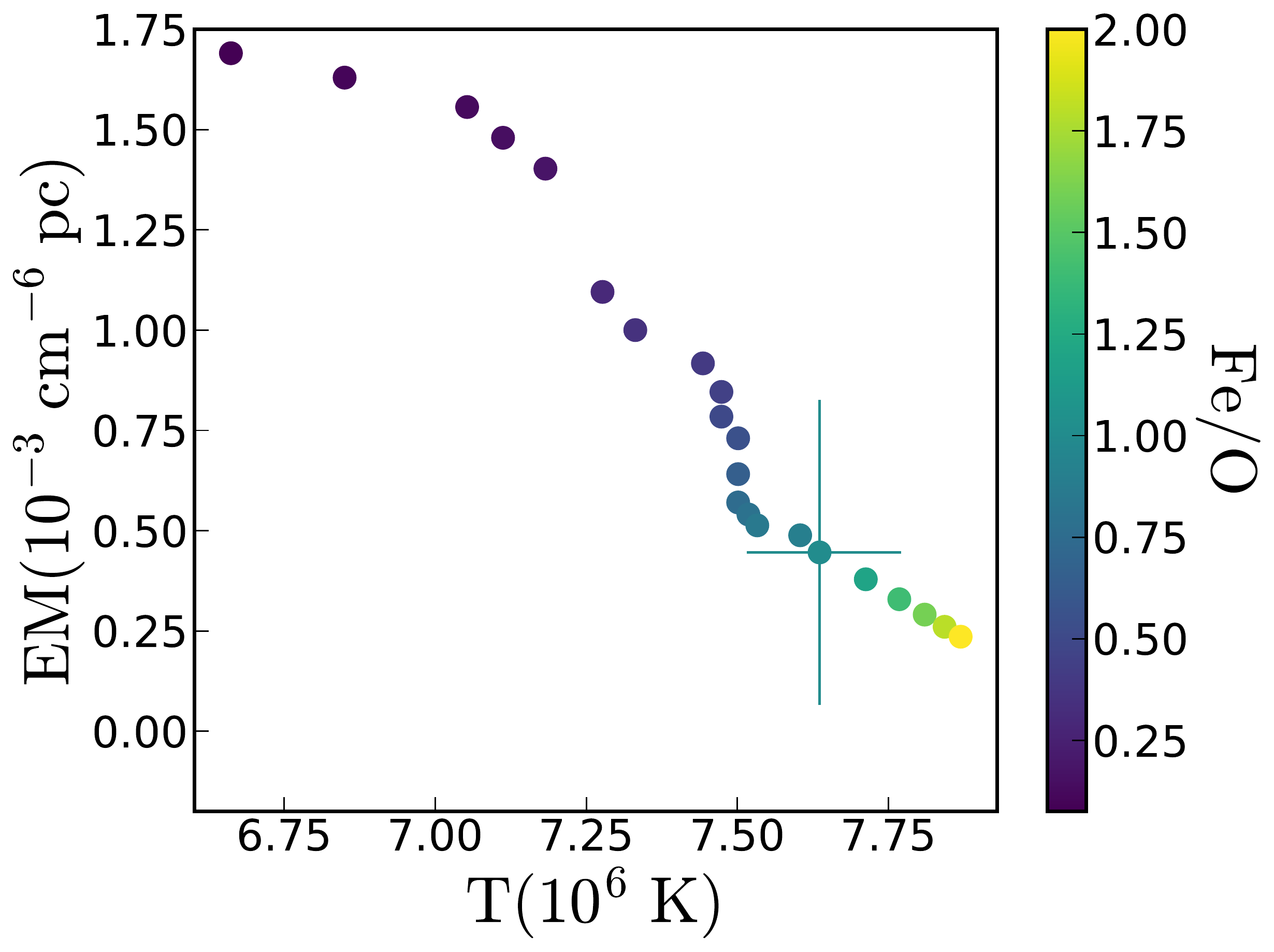}
\caption{The effect of varying the Fe abundance, $0.05<$Fe/O$<2.0$ on the best-fit estimate of the super-virial component in the T-EM parameter space. The error bar on the the point corresponding to solar abundance shows the typical uncertainty. Reducing the Fe abundance, leads to a higher EM and a lower temperature estimate. \label{Figure:Fe_abund}}
\end{figure}

 Another noteworthy result from  \citetalias{2021ApJ...918...83D} was the detection of $\alpha$-enhancement, with an upper bound on the Fe/O ratio of $0.22$ in the super-virial phase based on absorption. We tested the effect of varying the Fe/O levels on the EM and temperature measurements while keeping all other parameters frozen, using the MOS2 spectral data from the \mrk{} field. At Fe/O=0.22, we find EM$_2=1.78 \pm 0.30$ \emu{} and $kT_2 = 0.59_{-0.14}^{+0.10}$ keV. The results are shown in Figure \ref{Figure:Fe_abund}, where we see that a larger EM and lower temperature are predicted with increased $\alpha$-enhancement, i.e., lower Fe/O. This trend was observed in the auxiliary fields  with pn spectra as well. Thus we see that $\alpha$-enhancement would exacerbate the temperature differences between the emission and absorption measurements in the super-virial phase. 

\section{Discussion} \label{sec:discussion}

Our study of the fields around \mrk{} has led to several results on the CGM of the MW. In addition to the phase at the virial temeperature, the super-viral phase is clearly detected in all the fields. There is a range of observed temperatures, and EMs. The temperature measurement based on absorption is lower than that from emission in the virial phase. For the super-virial phase, on the other hand, the absorption based temperature is larger than those from emission. In the following we will try to understand these results. We will also investigate the relation between the virial and the super-virial phases, if any. Finally, we will comment on what we understand about the MW CGM based on our results. 

\subsection{The range of temperatures in the virial phase} \label{sec:emiss_bias}

As shown in Figure \ref{Figure:all_scatter} the temperature of the virial phase from emission study $T_{1,em}$ from all the fields (this work) lies beyond the 3$\sigma$ confidence interval of the temperature measured in absorption $T_{1,abs} = 1.5 \times 10^6$ K from \citetalias{2021ApJ...918...83D}. Here we infer the implications of this result. Not detecting absorption at $T_{1,em}$ implies a lower column density at $T_{1,em}$ compared to $T_{1,abs}$. 

Next let us understand why we do not detect emission at temperatures measured from absorption $T_{1,abs}$. This can result from a  combination of two parameters, emissivity and density. The dominant ion in both absorption and emission in this phase is O VII. 
The strength of absorption is driven by the O VII column density, which is proportional to the density as $N \propto n_{\rm O VII}(T)$.  The emission intensity, on the other hand, goes as $I \propto n_{\rm O VII}^2 (T) \epsilon (T)$ where the latter term represents the emissivity of the plasma at temperature $T$. The density in turn is proportional to the ionization fraction of O VII, $f_{\rm O VII}(T)$ which has a broad plateau over temperatures $10^{5.5} - 10^{6.3}$ K, thus absorption can probe a broad range of temperatures. On the other hand, emissivity  $\epsilon (T)$ peaks at $10^{6.3}$ K \citep{2009ApJ...690..143Y}, dropping sharply at lower and higher temperatures. For this reason, the emission spectrum is strongly sensitive temperature, probing the peak of $\epsilon (T)$. This explains why $T_{1,em}$ is larger than $T_{1,abs}$.

The temperature estimates based upon absorption and emission are weighed differently according to the line-of-sight plasma distribution with:

    \begin{equation}
    T_{em} = \frac{\int n(l)^2 T(l) dl}{\int n(l)^2 dl}, and\  
    T_{abs} = \frac{\int n(l) T(l) dl}{\int n(l) dl}
    \end{equation}

    Compared to $T_{abs}$, $T_{em}$ is weighed heavily from high density regions along the line-of-sight, which could be closer to the disk of the Galaxy. This would also result in $T_{abs} < T_{em}$, as observed for the virial phase. 

Thus we conclude that that there is no single temperature in the virial phase, but a true range of temperatures from $2.15\times 10^6 - 2.79\times 10^6$ K. The higher temperature range of this (observed in emission) has lower column density, and the low temperature (observed in absorption) has the higher column density. The emitting gas likely has higher density, possibly from regions close to the disk of the MW, while the absorption may arise from low-density gas extended out to the virial radius of the MW \citep{2012ApJ...756L...8G}; this is consistent with the conclusions of \citet{2020NatAs...4.1072K}.  The phase probed in absorption must be more massive than that probed by emission. 

\subsection{Range of temperatures in the super-virial phase} \label{sec:hot_how}

Figure \ref{Figure:all_scatter} shows that the  super-virial phase also has a range of temperatures $T_{2,em}$ from emission study (this work) and the temperature measured in absorption $T_{2,abs}$ is larger than $T_{2,em}$. In order to understand this, we need to investigate the characteristics of the emitting gas in more details. Our detection of the super-virial phase arises from Ne and Fe emission lines around 1 keV (see figures 3 and 4) \citep{1993ApJS...88..253S,2012ApJ...756..128F}. To understand the chemical makeup of the emitting gas, we present the relative contributions of these metals to the mock APEC spectra in Figure \ref{Figure:apec_spectrum_ONeFe} in Appendix \ref{sec:sv_iron}. We see that the Fe M-shell transitions at $\approx 0.75$ keV and especially the L-shell transitions at $\gtrsim 1.0$ keV play a more dominant role than the Ne IX and Ne X lines at $\approx 1.0$ keV, when the temperature of the gas is $>0.6$ keV.  In the bottom panel of Figure \ref{Figure:Feions_absem} we have plotted a figure of merit for emission, $f(T)^2 \times \epsilon (T)$. The dashed vertical lines mark the emission temperatures $T_{2,em}$ inferred for the five auxiliary fields as well as the \mrk{} field. We see that the range of observed $T_{2,em}$ comes from different ions from Fe XVIII--FeXXII dominating the emission through their L-shell transitions. 

The absorption temperature measurement  in \citetalias{2021ApJ...918...83D} is based on the Ne X and Si XIV ions, and Fe XVII-Fe XXV were not detected. In the top panel of Figure \ref{Figure:Feions_absem} we have plotted ionization fractions $f(T)$ of Fe XVII-Fe XXV.  Given that the ionization fractions of  Fe XVII-Fe XXV are negligible at $T_{2,abs}$, their non-detection is easily explained.

We conclude that there is a range of $T_{2,em}$ as probed by Fe XVIII--FeXXII emission. $T_{2,em}$ values are lower than $T_{2,abs}$ where the ionization fractions, and so the column density, of these ions is negligible. The fields around CL1115.8+4239 and 2MASXJ10514428 have $T_{2,em}$ within the 3$\sigma$ uncertainty of $T_{2,abs}$. Because of the chemical tagging, we can conclude that even these $T_{2,em}$ are distinct from $T_{2,abs}$, even though they may formally appear overlapping.

\begin{figure}
\centering
\includegraphics[scale=0.4,clip]{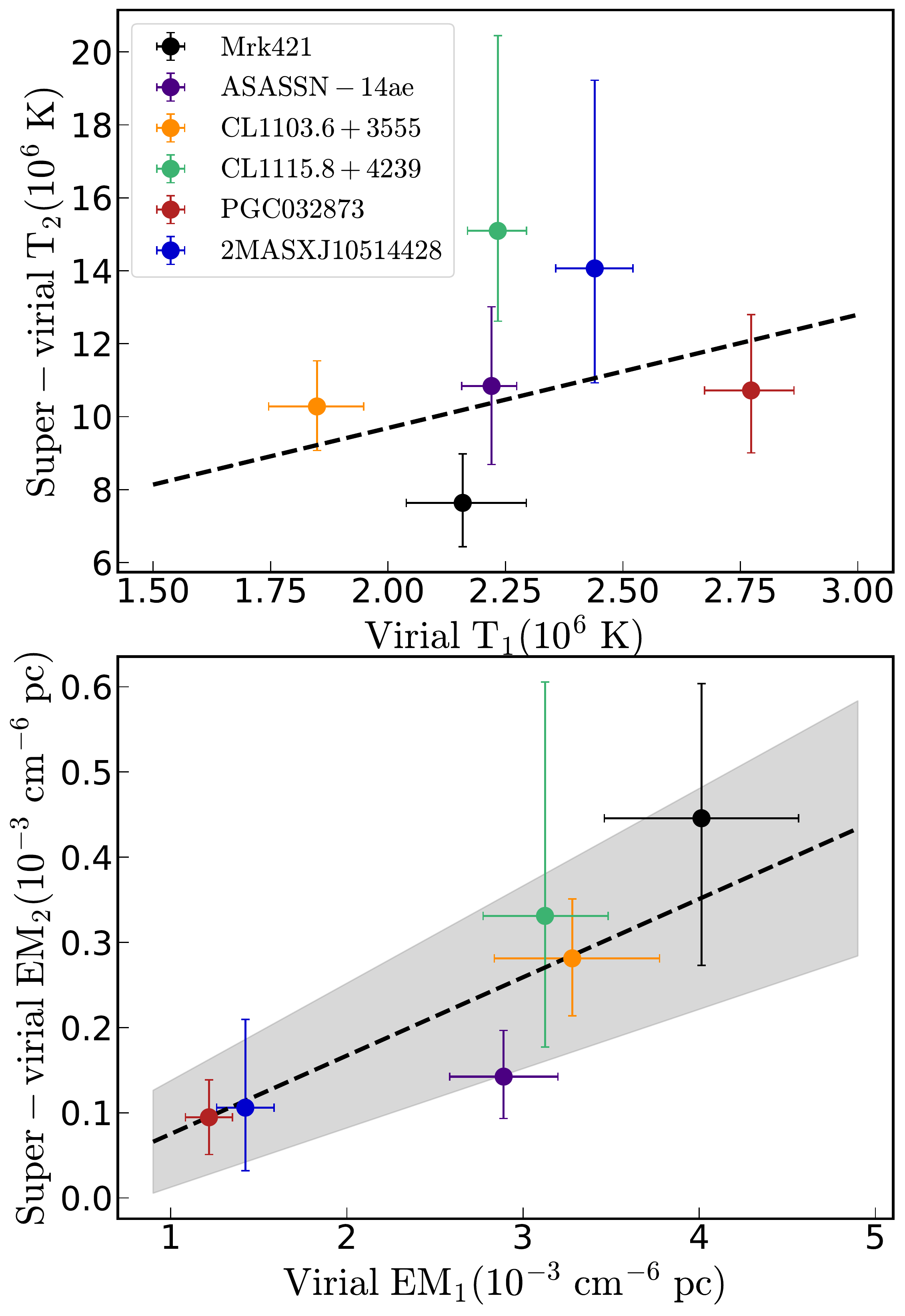}
\caption{ \textit{Top:} Temperature measurements from all the fields for the super-virial gas plotted against that of the virial gas. No correlation is observed. \textit{Bottom:} EM measurements for the super-virial gas plotted against that of the virial gas. The EM of the two phases appears to be strongly correlated. \label{Figure:param_corr}}

\end{figure}

\subsection{Anisotropy} \label{sec:patchy}

\citet{2009PASJ...61..805Y,2010ApJ...723..935H,2013ApJ...773...92H} have mapped the soft X-ray emission of the diffuse Galactic halo across the sky. They have all noted that while the temperature of the virial temperature gas is relatively uniform, the EM has a considerable scatter among the fields. The EM estimates can differ by as much as an order of magnitude across the sky. \citet{2020NatAs...4.1072K} observed numerous fields in the southern galactic sky ($b < -30$ deg) with HaloSat and found that the EM is patchy at scales of $\sim 10$ deg. Recently, \citet{2022arXiv220109915G} and \citet{2022arXiv220802477B} found patchiness across multiple fields, even in the super-virial gas while using a 2-T emission model. The patchiness can be attributed to spatially or temporarily stochastic star-formation or supernovae feedback \citep{2010ApJ...723..935H,2013ApJ...773...92H}.



In this work we have studied only a small patch of the sky ($\approx 80$ deg$^2$) with multiple fields within 5 degrees of the \mrk{} sightline. This allowed us to investigate the patchiness of the CGM at these small angular scales. We find that  (see Figure \ref{Figure:all_scatter}) the EM measurements of the virial temperature gas are scattered and have non-overlapping error bars, but the temperature estimates are relatively less dispersed. We calculate a fractional scatter to quantify this:

\begin{equation}
FS_T = \frac{\sqrt{\sigma_T^2 + \left< \delta_T^2 \right> }}{\left< T \right>}
\end{equation}

where $\sigma_T^2$ and $\left< \delta_T \right>$ represent the variance about the mean temperature and the mean of the squared temperature uncertainties, respectively. An equivalent formula is applied for the EM. We find that for the virial phase $FS_T = 14\%$ and $FS_{EM} = 44\%$, lending credence to our earlier observation. 

The super-virial component has large dispersions along both axes, with $FS_T = 34\%$ and $FS_{EM} = 80\%$. Whereas the scatter in the temperature measurements are linked to the nuances in emissivity discussed in Sections \ref{sec:emiss_bias} and \ref{sec:hot_how}, the dispersion in EM implies that the super-virial component is patchy on these angular scales, and also is more patchy compared to the virial component. This might be because of the intrinsic anisotropy of density, temperature, metallicity, and/or abundances of these phases. Theoretical simulations \citep{2015MNRAS.446..521S,2016MNRAS.460.2157O} predict density and metallicity inhomogeneity for a MW-like galaxy.
Similarly, simulations of \citet{2016ApJ...827..148C} show that the CGM of
a MW-like galaxy has significant filamentary structure. Our observations of inhomogeneity are broadly consistent with these predictions, though the predicted angular scales are unclear from the papers.  We conclude that MW CGM is inhomogeneous even on small scales of a few degrees, 
 because of clumpiness and/or filamentary structure,
caused by physical processes such as turbulence, outflows, small- and large-scale environment. 

\subsection{Relation between the virial and super-virial phases} \label{sec:correlation}

The MW CGM was known to have a warm-hot phase at the virial temperature. Is there any relation between the newly discovered super-virial phase and the diffuse, extended, massive, virial phase? We investigate this, we search for correlations in the best-fit parameters of all the fields
in Figure \ref{Figure:param_corr}. In the top panel, we plot the temperatures of the super-virial and virial components along the two axes. There is a scatter without any noticeable correlation between these two parameters. Nonetheless, we fit a straight line to these points using Orthogonal Distance Regression \citep{joint1990statistical} to get a slope of $1.14 \pm 2.95$ and a Pearson correlation coefficient of 0.129. We repeat the same exercise for the EM measurements in the lower panel, where they show a relatively stronger correlation between the super-virial and virial phases. The slope of the best-fit straight line is found to be $0.094 \pm 0.024$, and a correlation coefficient of 0.729 may hint at an underlying relationship.
Interestingly, the slope and correlation coefficient between the EM of the two components agree exactly with the results of \citet{2022arXiv220802477B} using multiple fields across the sky. 

This correlation may also result from clumpiness, or density inhomogeneity. The fields with higher density would have higher EM at both $T_{1,em}$ and $T_{2,em}$ compared to fields with lower density. 

\subsubsection{Pressure balance}

\begin{figure}
\centering
\includegraphics[scale=0.35,clip]{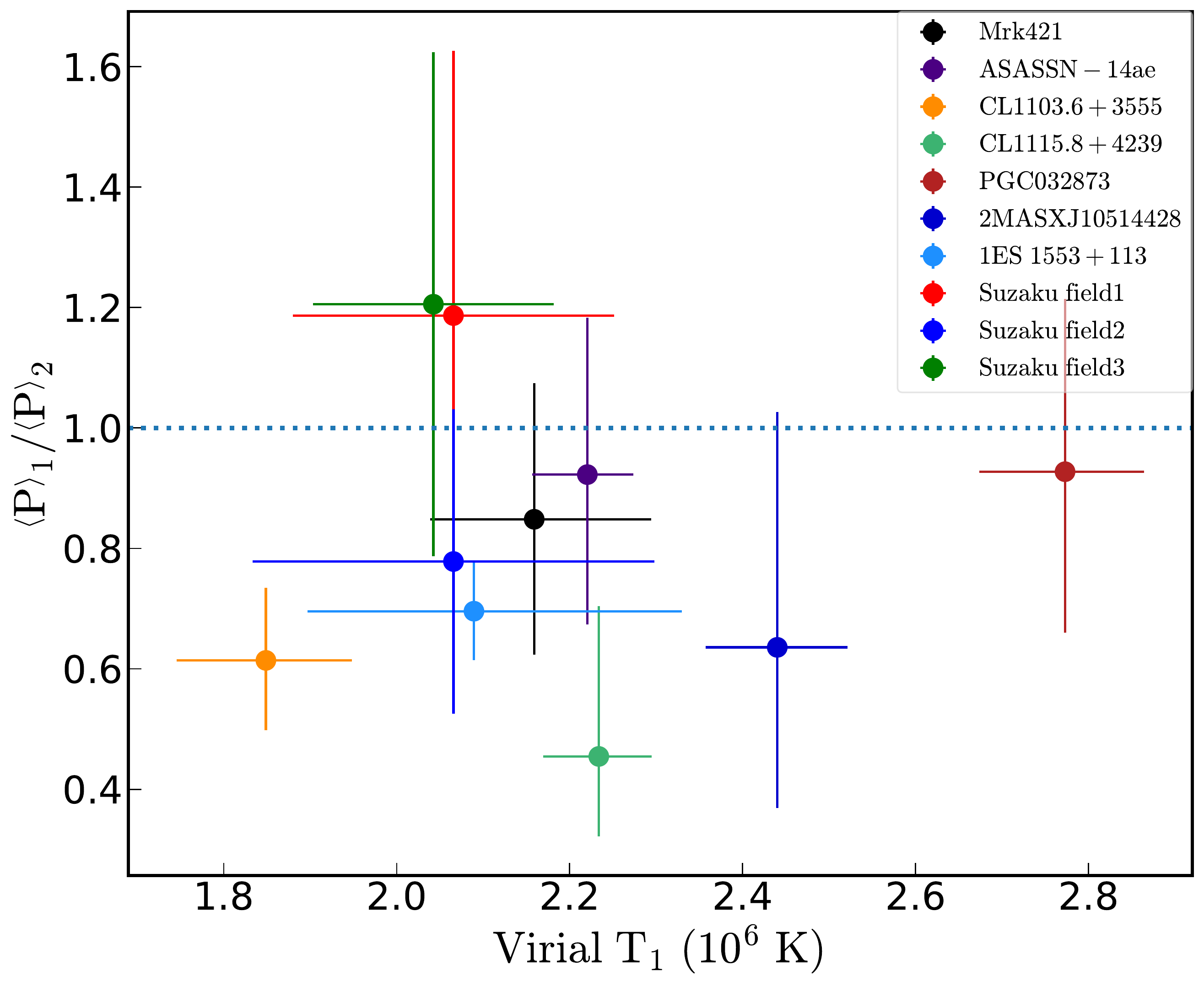}
\caption{ The ratio of average pressures of the virial gas with respect to the super-virial among the various sight lines plotted against the temperature of the former. Estimate using the values from \citet{2019ApJ...887..257D,2021ApJ...909..164G} have also been plotted. There is a general deviation from prssure equilibrium with the virial gas being at a lower pressure. \label{Figure:press_scat}}

\end{figure}

In Figure \ref{Figure:all_scatter} we see that the lower temperature virial phase has higher EM, and the higher temperature super-virial phase has lower EM. 
In order to test whether these two phases of the CGM exist in pressure equilibrium, we calculate the ratio of the average pressure of the virial and super-virial phases $\left< P \right>_1$ and $\left< P \right>_2$ respectively. We assume that gas is uniformly distributed such that, EM=$\left< n^2 \right> L$, with $L$ representing the path-length. The average density can then be derived as $\left< n \right> = \sqrt{EM/L}$. Assuming the gas is in thermal equilibrium, the average pressure will be $\left< P \right> = \left< n \right> T$. The ratio of pressures for the two components is:

\begin{equation}
\frac{\left< P \right>_1}{\left< P \right>_2} = \frac{\sqrt{\left< EM_1\right>} T_1}{\sqrt{\left< EM_2 \right>} T_2}  
\end{equation}

assuming that both phases have the same $L$. In Figure \ref{Figure:press_scat} we plot the ratios calculated for the fields in this work, as well as using the results of \citet{2019ApJ...887..257D,2021ApJ...909..164G}. We find that $\left< P \right>_1/\left< P \right>_2 < 1$ for 7 fields, while 5 of their error bars overlap with the $\left< P \right>_1 = \left< P \right>_2$ line. Two points have $\left< P \right>_1/\left< P \right>_2 < 1$ but they have  large uncertianties. Given the distributions in Figure \ref{Figure:press_scat}, we cannot conclude that the virial and super-virial phases are in pressure balance with each other. Moreover, as discussed above, there is no single virial temperature or a single super-virial temperature. The CGM has a range of temperatures even in the range probed in X-rays. A simple pressure equilibrium between two phases is unlikely. 

\subsection{Location of the super-virial phase}

The super-virial phase of the MW CGM is newly discovered, and we do not yet know its location within the Galaxy, or its origin.  The presence of this component \citep{2019ApJ...887..257D,2022arXiv220109915G} around the Galactic bubbles \citep{2020Natur.588..227P,2010ApJ...724.1044S} has been tied to relics of outflows that caused these structures, e.g., historic AGN \citep{2020ApJ...894..117Z,2022NatAs...6..584Y} or nuclear star-formation activity \citep{2015MNRAS.453.3827S,2017MNRAS.467.3544S}. On the basis of the size of the bubbles and the gas temperature, \citet{2022arXiv220109915G} argue in favor of isothermal shocks driven by nuclear star-formation activity as the cause for the bubbles. However, the \mrk{} sightline ($l=179.83 \deg,b=+65.03 \deg$)  is away from the regions of the Galactic bubbles, or the nucleus. Hence, the super-virial gas in the fields around \mrk{} arose from a process (or processes) different from what formed the bubbles. 

The super-virial gas can be attributed to an extra-galactic hot reservoir of gas, e.g., the intergalactic medium (IGM) \citep{1999ApJ...514....1C,2001ApJ...552..473D} which is located beyond the halo of the Milky-Way. However, it is unlikely that the gas is associated with the Local Group medium since the sight lines are located away from the location of M31 or its barycenter \citepalias{2021ApJ...918...83D}. We concluded in Sections \ref{sec:emiss_bias} and \ref{sec:correlation} that the phases observed in emission likely arise from denser gas closer to the Galaxy disk. If this is the case, it may be a result of hot gas outflow \citep{2008MNRAS.386.1464R,2020ApJ...890L..30L,2020ApJ...898..148L,2020ApJ...904..152L}. 

\section{Conclusion} \label{sec:conclusion}

We have used archival observations made using the EPIC instrument onboard XMM-Newton to study the soft X-ray emission of the CGM of the Milky-Way. We chose the field around  \mrk{} because there has been a complementary study in absorption \citepalias{2021ApJ...918...83D} and also due to the fact that there exists a large amount of archival data for this field.  We further investigate the soft X-ray emission of the CGM in 5 auxiliary fields located within five degrees of the \mrk{} sightline. Our main results are listed below.

\begin{enumerate}
    \item In the field around \mrk{} we detect the presence of a super-virial hot component of the CGM with log ($T/$K) = $6.88_{-0.07}^{+0.08}$ and EM=$0.45_{-0.17}^{+0.15}$ \emu{} in addition to the virial component with log ($T/$K) = $6.33_{-0.02}^{+0.03}$ and EM=$4.01_{-0.55}^{+0.54}$ \emu{}.   
    \item The detection of super-virial component is also confirmed in 5 other fields within 5 deg of \mrk{}. From a simulteneous fit to the all of the spectra we get a temperature and emission-measure of the super-virial phase as log ($T/$K) = $7.08_{-0.03}^{+0.04}$ and EM=$0.14\pm0.03$ \emu{}. 
    \item We perform chemical tagging to conclude that the emission from the super-virial phase is produced  by the Fe L-shell transitions of Fe XVII-Fe XXV; the observed range of temperatures in emission is distinct from the temperature measure in absorption probed by Ne X and Si XIV. 
    \item A model with both virial and super-virial phases having solar abundances is found to fit the spectra better than a single component virial gas with a Neon over-abundance (Ne/O$\approx$2), resolving the previous degeneracy in results. 
    \item The emission is patchy on scales of a few degrees. 
    \item We find a correlation between the emission-measures of the virial and super-virial components which is stronger than any correlation between the temperatures of the two components. 
    \item The fractional scatter of both the emission-measures and temperatures across all fields are found to be larger for the super-virial gas compared to the virial gas.
\end{enumerate}

With our results, we draw interesting conclusions about the MW CGM. Even in the range probed in X-rays, the CGM has a range of temperatures in the virial phase around $2 \times 10^6$ K and in the supervirial phase around  $10^7$ K. Since we probe only a small patch of the sky, any single density and/or temperature profile cannot account for the observed scatter in temperature and EM; there has to be small-scale density inhomogeneity instead.  This may be caused by physical processes such as turbulence, outflows, small-scale environment, as suggested by theoretical simulations. 
The emitting gas likely has higher density, possibly from regions close to the disk of the MW, while the absorption in the virial phase may arise from low-density gas extended out to the virial radius of the MW \citep{2012ApJ...756L...8G}.  Therefore, the phase probed in absorption in the virial phase must be more massive than that probed by emission. The presence of the super-virial phase far from the regions around the Galactic center implicates physical processes unrelated to the activity at the Galactic center. Hot outflows resulting from star-formation activity throughout the Galactic disk are likely responsible for producing this phase. We speculate that the gas detected in emission has cooled from the hot outflow detected in absorption. 

\acknowledgments

SB received support from the Ohio State University Graduate Fellowship. AG and SM are grateful for the support through the NASA ADAP grants NSSCK0480 and 80NSSC22K1121, respectively. 

\software{XMM SAS \citep{2004ASPC..314..759G}, XSPEC \citep{1999ascl.soft10005A}, Astropy \citep{2013A&A...558A..33A,2018AJ....156..123A},
DS9 \citep{2000ascl.soft03002S}, pyAtomDB \citep{2020Atoms...8...49F}, CHIANTI \citep{1997A&AS..125..149D} }

\bibliography{sample63}{}
\bibliographystyle{aasjournal}

\appendix

\section{Elemental makeup of the super-virial gas} \label{sec:sv_iron}

\begin{figure*}[h!]
\centering
\includegraphics[scale=0.35,clip]{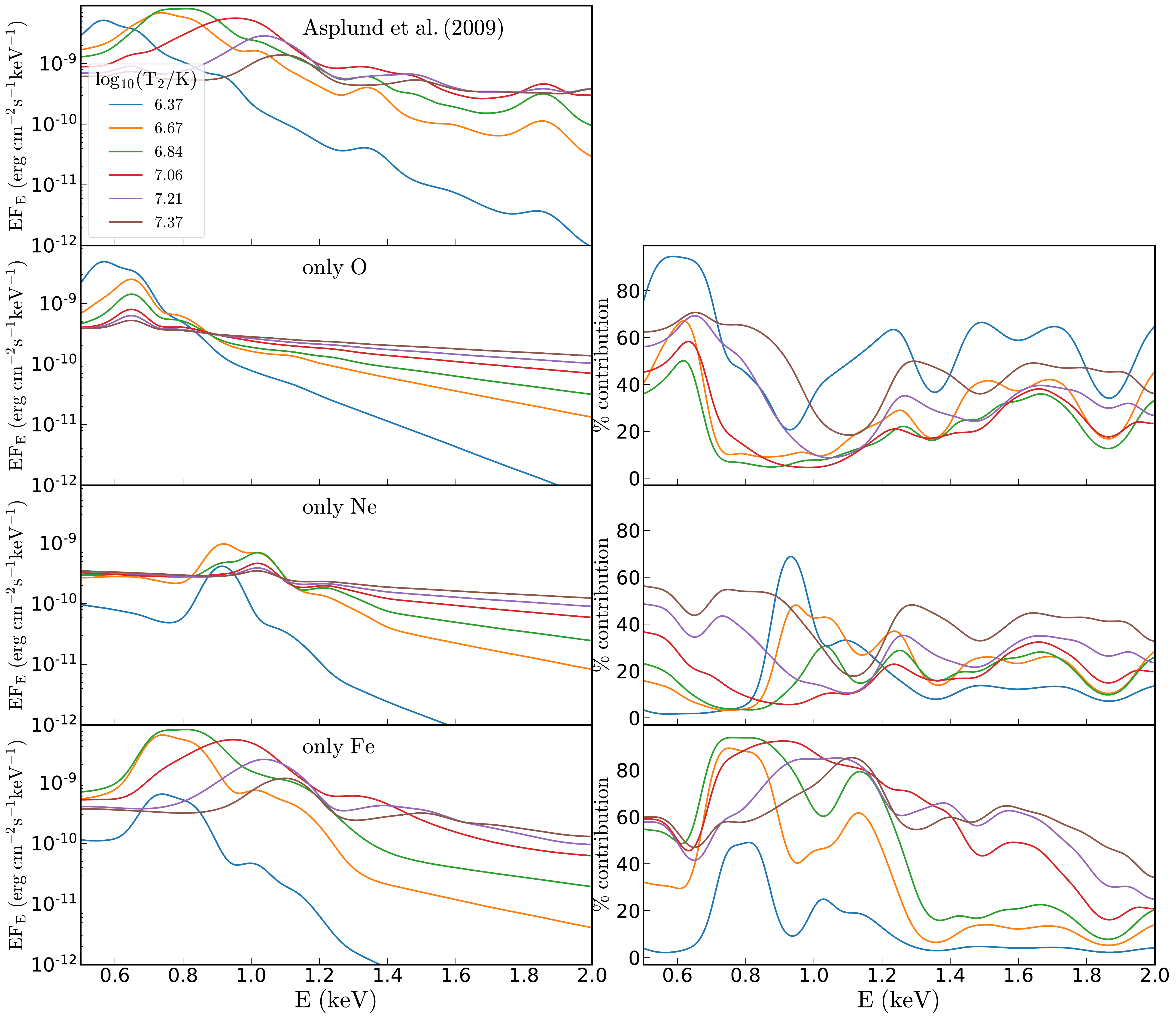}
\caption{\textit{Left:} The SEDs for plasmas with a range of temperatures 0.2 $<kT<$ 2.0 keV simulated using APEC, for different elemental abundances. The topmost row is for solar composition \citep{2009ARA&A..47..481A} while the subsequent rows depict the SEDs where each of O, Ne and Fe are the only elements present. \textit{Right:} The ratios of the O, Ne and Fe only SEDs are taken with respect to the SED with solar composition, to show how each of the elements contribute to the latter at different energies and $kT$.}
\label{Figure:apec_spectrum_ONeFe}
\end{figure*}

We use pyAtomDB \citep{2020Atoms...8...49F} to study the elements and their corresponding ionic species mainly contributing to the emission spectra of the super-virial phase at energies of $\sim$ 1 keV. For this purpose we simulate the APEC spectra at different plasma temperatures $kT>0.6$ keV
and fold the model with the response matrix of the EPIC-pn camera. The SEDs with contributions from only O, Ne and Fe are compared with the SED from a gas with solar composition \citep{2009ARA&A..47..481A} in Figure \ref{Figure:apec_spectrum_ONeFe}. At $kT =0.2$ keV, we notice the prominence of the O VIII and Ne IX lines at 0.6 keV and 0.9 keV respectively, the latter of which was the possible source of degeneracy, which we addressed in Section \ref{sec:abundances}. At higher temperatures of $kT\sim 0.6$ keV, we find that the Fe L-shell transitions play a far dominant role, with there being a range of peaks at progressively higher energies for different $kT$. They arise mainly from the ionization states of Fe XVII and higher \citep{1993ApJS...88..253S,2012ApJ...756..128F}; we investigate this further in Section \ref{sec:ion_iron} below.

\section{Fe ionization states in the super-virial gas} \label{sec:ion_iron}

In Appendix \ref{sec:sv_iron}, we saw that the Fe L-shell transitions play a dominant role in the soft X-ray emission for a super-virial gas at $\sim$ 1 keV. Here we determine, for each of the various ionization states of Fe, their contribution to the absorption and emission spectra. Alongwith data from AtomDB, we also use data for the ionization equilibria for these states from the CHIANTI database \citep{1997A&AS..125..149D}.

\begin{figure*}[h!]
\centering
\includegraphics[scale=0.35,clip]{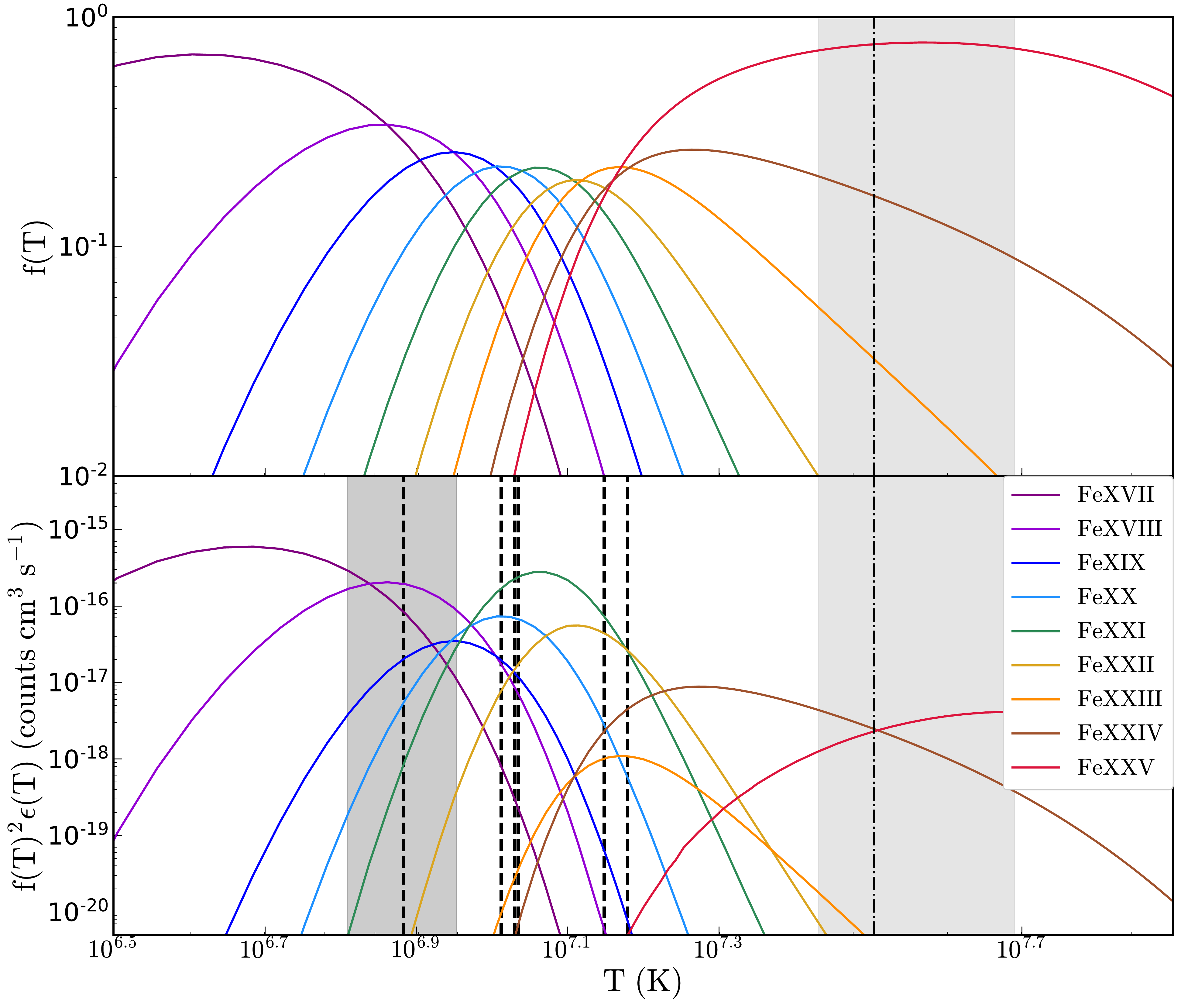}
\caption{\textit{Top:} The strength of absorption signified by the ionization fractions $f(T)$ as a function of temperature for different ionization states Fe XVII - Fe XXV represented by the different colored lines. \textit{Bottom:} The strength of emission signified by the product of emissivity $\epsilon (T)$ with the square of $f(T)$ as a function of temperature for the same ionization states. The \textit{dash-dotted} lines with the gray shaded region shows the temperature estimate of the super-virial gas from \citetalias{2021ApJ...918...83D} and the associated $2\sigma$ uncertainty. The \textit{dashed} lines are the measured temperatures from the different sightlines shown in \ref{Table:PNenvironlist} with the shaded region showing the uncertainty in the estimate along the \mrk{} sightline.  }
\label{Figure:Feions_absem}
\end{figure*}

The upper panel of Figure \ref{Figure:Feions_absem} shows the ionization fractions $f(T)$ as a function of temperature, with the lines of different colors representing the various ionization states Fe XVII-Fe XXV. This is related to the strength of absorption at different temperatures by these ions. From the lower to higher ionization states, the peaks in their $f(T)$ occur at successively higher temperatures. However we notice that Fe XVII and Fe XXV reach a relative higher maxima compared to the other states. In the lower panel we show a figure-of-merit for emission, as the product of emissivity $\epsilon (T)$ and the square of $f(T)$ as a function of temperature. We notice peaks in the range $10^{6.7} - 10^{7.8}$ K. 
The emission and absorption based measurements of the temperature of the super-virial gas from this work and \citetalias{2021ApJ...918...83D} have been overlaid. We can see that emission is primarily dictated by the Fe XVIII-Fe XXII series of states.

\end{document}